\documentclass[conference,twocolumn]{IEEEtran}
\IEEEoverridecommandlockouts
\usepackage{cite}
\usepackage{amsmath,amssymb,amsfonts}
\usepackage{algorithmic}
\usepackage{graphicx}
\usepackage{textcomp}
\usepackage{xcolor}
\usepackage{subcaption}

\usepackage{tikz}
\usepackage[utf8]{inputenc}
\usepackage{siunitx}
\usepackage{pgfplots}

\newcommand\orcidicon[1]{\href{https://orcid.org/#1}{
\includegraphics[height=0.8\baselineskip]{orcid.ps}
}}

\DeclareUnicodeCharacter{2212}{−}
\usepgfplotslibrary{groupplots,dateplot,polar,external}
\usetikzlibrary{patterns,shapes.arrows}
\pgfplotsset{compat=newest,every axis/.append style={font=\scriptsize,
                              width=\linewidth,
                              height=0.6\linewidth}}

\def\BibTeX{{\rm B\kern-.05em{\sc i\kern-.025em b}\kern-.08em
    T\kern-.1667em\lower.7ex\hbox{E}\kern-.125emX}}

\begin{document}

\title{3D beamforming and handover analysis for UAV networks\\
\thanks{This research is supported by the Research Foundation Flanders (FWO), project no. S003817N (OmniDrone) and by the European Union’s Horizon
2020 under grant agreement no. 732174 (ORCA project) .}
}

\author{\IEEEauthorblockN{
Achiel Colpaert, %\orcidicon{0000-0003-1238-2002},
Evgenii Vinogradov, %\orcidicon{0000-0002-4156-0317},
and Sofie Pollin %\orcidicon{0000-0002-1470-2076}
}
\IEEEauthorblockA{KU Leuven, ESAT - Department of Electrical Engineering, Kasteelpark Arenberg 10, 3001 Heverlee, Belgium\\
E-mail: \{achiel.colpaert, evgenii.vinogradov, sofie.pollin\}@kuleuven.be}}

\maketitle

\begin{abstract}
In future drone applications fast moving unmanned aerial vehicles (UAVs) will need to be connected via a high throughput ultra reliable wireless link. MmWave communication is assumed to be a promising technology for UAV communication, as the narrow beams cause little interference to and from the ground. A challenge for such networks is the beamforming requirement, and the fact that frequent handovers are required as the cells are small. In the UAV communication research community, mobility and especially handovers are often neglected, however when considering beamforming, antenna array sizes start to matter and the effect of azimuth and elevation should be studied, especially their impact on handover rate and outage capacity. This paper aims to fill some of this knowledge gap and to shed some light on the existing problems. This work will analyse the performance of 3D beamforming and handovers for UAV networks through a case study of a realistic 5G deployment using mmWave. We will look at the performance of a UAV flying over a city utilizing a beamformed mmWave link.
% We will investigate the handover situation in both a 4G deployment without beamforming and a 5G deployment with beamforming. We suggest improvements for the handover thresholds in both scenario's.
% Secondly we will investigate the effect of these handovers on the reliability of the system. Can we improve this reliability further by considering packet duplication/dual connectivity?
\end{abstract}


\section{Introduction}
\subsection{Background}
\par Drones are becoming more and more ubiquitous in our daily life. Several industries are looking at drones to improve or create new services. In these new applications unmanned aerial vehicle (UAVs) often operate in a beyond-line-of-sight manner and to control these applications a reliable wireless link is required albeit with limited capacity. However, some companies have expressed their interest in streaming high definition (HD) video over a wireless link from a drone requiring both a high throughput and reliable link.
\par Serving UAVs in beyond-line-of-sight operation by using the existing cellular network is the next logical step. Cellular networks provide omnipresent coverage and they should support the requirements for command-and-control and even high throughput links. The 3GPP initiative has even taken the first steps towards incorporating UAV users into their future standards by publishing technical report on this topic \cite{3gpp.36.777}.
\subsection{State of the art}
\par While these cellular networks sound promising, research has indicated that introducing drones into cellular networks will create several problems. First of all, authors of \cite{ Colpaert2018, Azari2018} show using either stochastic or semi-deterministic simulations that aerial users experience mainly line-of-sight (LOS) propagation conditions to multiple basestations (BS). While this is an advantage when considering the signal of the serving BS, it has been proven problematic in terms of interference caused by neighbouring BSs. They show that UAVs operate mainly in the interference limited domain instead of the usual noise limited domain for ground users. Other works have confirmed these results by performing outdoor experiments \cite{QualcommTechnologies2017, VanDerBergh2016}. The first found that the link quality at several altitudes should be sufficient to support command-and-control operation, although they mainly performed measurements in a semi-urban, close to rural, environment \cite{QualcommTechnologies2017}. The second showed through measurements at larger altitudes that the number of BSs causing interference rises dramatically when the UAV rises in altitude \cite{VanDerBergh2016}.
\par Another problem introduced by aerial users is the interference generated in the uplink to other BSs \cite{VanDerBergh2016}, as well as the interference to to other ground applications such as satellite ground stations \cite{Vinogradov2019}. This last work suggests some interference countering techniques such as simply defining no-fly zones, more advanced antennas on the UAV side or the use of beamsteering.
\par Several solutions have already been proposed to solve these interference issues. Optimizing the cellular network configurations has been suggested by \cite{Azari2019}, such as optimizing the BS antenna tilt, however we cannot assume that providers will sacrifice performance for ground users in order to support aerial users. Several works suggest that directive antennas and beamforming will allow the BS to spatially separate users in 3D space allowing for efficient serving of both ground and aerial users. In this context, researchers mention that 5G is a good candidate as it supports beamforming and high throughput links, while being highly configurable \cite{Lin2018}. On the other side, authors of \cite{Izydorczyk2020} suggest to put highly directive antennas on the UAV. Another solution is proposed by the authors of \cite{Garcia-Rodriguez2019}. They suggest to use the massive MIMO support in 5G which allows the BS to not only spatially separate the the users but also create nulls at other users to further reduce the interference. In previous work, we proposed mmWave communications as a promising technology for drones as they experience mainly LOS propagation, which is a mayor requirement for mmWave communications to work \cite{Colpaert2018}. The larger pathloss at mmWave frequencies would result in less inter-cell interference, while the small antenna aperture size allows for the use of large number of antennas in an antenna array. These large antenna arrays are ideal to perform beamforming to compensate for the large pathloss at the user while simultaneously reducing interference. The use of mmWaves opens a wide spectrum to use and the high throughput mentioned earlier can be easily achieved by utilizing massive bandwidth.
\par Previously mentioned research considered mainly static users, however, when mobile scenarios are considered even more problems arise. First of all, beam training and tracking becomes more difficult and generates a lot of overhead, however, this overhead is less than initially thought in research and thus should be able to support mobile users up to decent speeds \cite{Huang2020}. Another problem with using mmWaves is large Doppler frequency shifts as these are proportional to the center frequency. Authors of \cite{Stanczak2018} summarize several problems that arise for mobile aerial users. Aerial users are most of the time served by sidelobes of the antenna pattern in current static cellular deployments. This results in a fragmented association pattern. This fragmented association in combination with the low signal-to-interference-noise-ratio (SINR) results in a higher probability of radio link failures as well as handover failures. Due to the fragmented association pattern more ping-pong handovers will take place, where a user is handed over back to its original cell within a certain time frame. While LTE is designed to support users traveling at speeds up to 350~km/h\cite{3gpp.36.101} it assumes large cell areas and not these sidelobe based cell association patterns that UAV's experience. The 3GPP study item has identified cell selection, handover efficiency and robustness as a key performance indicators for aerial users in cellular networks\cite{3gpp.36.777}.
\par Euler et al. \cite{Euler2019} investigate this problem further on a simulation basis and they point at two problems. First, they conclude that high interference levels will make it difficult to maintain connection as well as perform successful handovers, which will lead to a large number of both radio link failures as well as handover failures.They explore is the usage of LTE-M which allows users to operate in low SINR conditions. They were able to significantly reduce the number of radio link failures and handover failures with only a slight increase in ping-pong handovers.
Secondly, they conclude that when aerial users move through antenna pattern nulls the default handover mechanism will be too slow to prevent radio link failure. To solve this issue, they suggest to tune the parameters of the handover procedure such as the reaction time.
\par With the deployment of 5G networks, \cite{Muzaffar2020} performed some preliminary experiments with a drone connected to a 5G BS at sub-6~GHz frequency. They concluded that the UAV experienced more handovers than a ground user but also that handovers to the 4G network occurred regularly reducing the overall throughput, but this will be solved by the deployment of more 5G BSs.
\par Aside from previous mentioned research, handover problems are largely neglected in the UAV research community when investigating cellular connected UAVs, especially in the context of beamforming. In the same trend, the effect of the used antenna array topology when performing beamforming is also largely disregarded in UAV studies, while in our opinion the antenna topologies can have a big influence on the behaviour of beamsteering and handovers alike.
\subsection{Problem statement}
\par In this work, we fill some of this knowledge gap regarding the effect of antenna topologies and the effect of beamforming on handover problems. This work poses as a stepping stone in the path towards the design of optimal beamforming networks for UAV communications. Regarding the antenna array topology, we investigate the effect using different topologies of different sizes and their effect on the handover behaviour and outage cost. Especially in the context of mmWaves, where a single antenna element has a physical size in the order of magnitude of millimeters, hence large number of mmWave antennas can be manufactured in a physically small antenna array \cite{Rappaport2013, Colpaert2020}. These large number of antennas allow us to create large antenna gains through beamforming, however, at the same time when using beamforming with large antenna arrays, the beamwidth becomes more and more narrow with increasing array size. One can see that at a certain point the difficulty of tracking the user with a beam as wide as a pencil will counter the gains won of using more antenna elements. Initial access will  pose problems in terms of beam discovery, although several tricks can be applied where initially wider beams are used to locate the user. When introducing user mobility even more problems arise. Keeping track of the user device with almost no margin for error proves a difficult task, especially if this user device is a UAV or any other device that needs a reliable connection to the network and is moving at larger speed than walking speed.
\subsection{Contributions}
\par This work will show the main issues that arise when a UAV is flying over a semi-urban area while being served by 5G mmWave BSs on the ground. We will look at the handover problems and the effect of beam misalignment and how choosing the correct antenna topology could counter these problems. The main contributions of this work can be listed as follows:
\begin{itemize}
    \item Provide insight in the effect of increasing the number of antenna elements on the beamwidth and the outage probability at the UAV side;
    \item Stress the importance of highly accurate beamtracking with increasing size of the antenna arrays;
    \item Show that the UAV operating altitude should be considered when designing the antenna array on BS side;
    \item Indicate the effect of choosing different antenna array topologies on the number of handovers experienced by a UAV.
\end{itemize}
The rest of this paper is organized as follows. First, we shortly discuss the simulator environment used to simulate the behaviour of the UAVs and the network. Followed by the additions implemented to support the necessary beamforming and handovers techniques for the scenarios. Next we discuss the investigated scenarios and the parameters used. Then, we show and discuss the results and, lastly, we conclude our work.
% \sofie{or add a figure where you show the problem statement/contributions clearly to make it more clear what the focus of the paper is?}

\section{Methods}
\par
\begin{figure}[tbp]
\centering
\centerline{\includegraphics[width=1\linewidth]{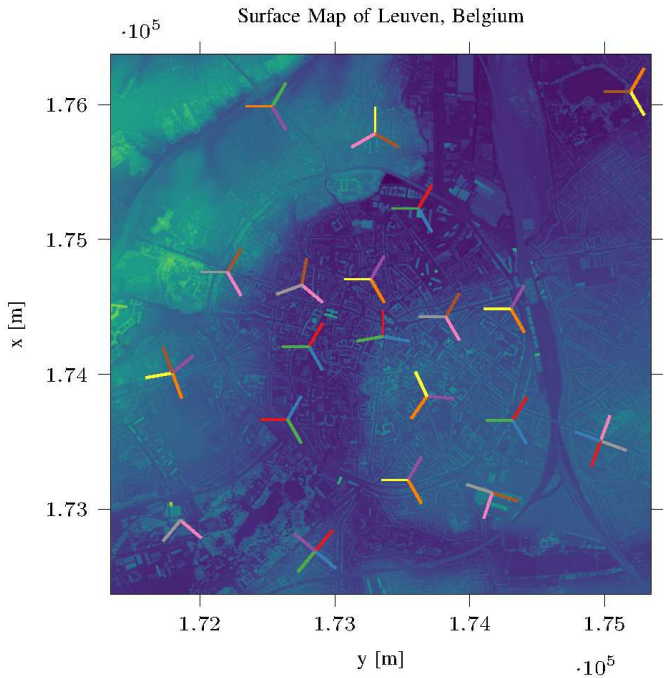}}
\caption{Representation of the considered area where each pixel represents the altitude above sea-level. All BS locations are indicated with their respective sectors and orientation.}
\label{fig:area}
\end{figure}

 To simulate the behaviour of the UAV flying over the city we use the coverage simulator described in \cite{Colpaert2018}.
It applies the 3GPP wireless channel models \cite{3gpp.38.901} in a 3D environment to determine the received signal strength at the users. We assume a 5G deployment where a current internet service provider's (ISP's) deployment sites are equipped with 5G mmWave capabilities without adding extra locations. The location considered is the city of Leuven in Belgium and the BS deployment of a local ISP is used, where each BS has either three or four sectors each. The considered area can be seen in Figure~\ref{fig:area}.
%The x and y-axis represent Lambert coordinates which is a localisation system used within Belgium where one pixel represents one square meter \cite{lambert}.
In terms of a mmWave deployment this should be a sub-optimal configuration for ground users and further cell densification is necessary to provide decent city-wide coverage. However, for UAVs, this poses less of a problem as UAVs experience almost always LOS connections to their serving BSs, which is the ideal case for a mmWave connection.
\subsection{Beamforming and handover simulations}
Two new features have been introduced in previously mentioned simulator. First of all, support for antenna radiation patterns of antenna arrays is implemented, including the use of steering vectors. This allows for advanced beamforming and beamtracking simulations. For the single element of the array, the patch antenna pattern is used as defined in \cite{3gpp.38.901}. To simulate the exact antenna array pattern, we use the array factor defined in antenna theory. The antenna array gain in a direction is defined in \cite{Balanis2016} as $G(\theta,\phi) = G_0(\theta,\phi) * AF,$ where $\theta$ represents the elevation angle measured from the z-axis which is defined as the zenith, $\phi$ is the azimuth angle in the horizontal plane with north defined as 0 degrees, $G_0(\theta,\phi)$ the single element gain in given direction and where $AF$ represents the array factor defined as $AF = S_{z_M} \cdot S_{y_N}$,
% \begin{equation}
%     AF = S_{z_M} \cdot S_{y_N},
% \end{equation}
where:
\begin{equation}
    S_{z_M} = \sum_{m = 1}^{M}I_{m1}e^{j(m-1)(kd_zcos(\theta)+\beta_z)}
\end{equation}
\begin{equation}
    S_{y_N} = \sum_{n = 1}^{N}I_{1n}e^{j(n-1)(kd_ysin(\theta)sin(\phi)+\beta_y)},
\end{equation}
where $M$ and $N$ represent the number of antenna elements on the z and y-axis of the array, respectively, $k$ is the wave number, $d_z$ and $d_y$ are the distances between the elements on z and y-axis respectively and where $\beta_z$ and $\beta_y$ represent the factors introduced by the steering directions and they can be calculated as follows:
\begin{equation}
\beta_z = -kd_zcos(\theta_0)
\end{equation}
\begin{equation}
\beta_y = -kd_ysin(\theta_0)sin(\phi_0)
\end{equation}
where $\theta_0$ and $\phi_0$ represent the desired beamsteering direction. When discussing different array topologies we will always mention two numbers, the first being the number of antenna elements in the vertical domain $M$ and the second being the number of elements in the horizontal domain $N$. To achieve %an apples-to-apples
a fair comparison, we keep the number of antenna elements constant when comparing different antenna array shapes
%, eg. 1 by 256 compared to 128 by 2.
\par %A second new addition to the simulator are mobile users.
Since in this paper, we aim to investigate the beam tracking effect on handovers, we have added mobility support to our simulator. While previously all users were static and snapshots in time were taken and processed, now users can move around in space and their parameters can be simultaneously monitored. Next to this mobility, a basic messaging framework was implemented to support handovers. In this framework, a user will monitor its A3 event such as defined in the 5G standard \cite{3gpp.38.331}. This event triggers when a neighbouring cell signal becomes stronger than the serving cell with a certain threshold. The threshold for this event is the A3 threshold ($T_{A3}$) and is defined in Table \ref{tab:parameters}. When the event requirements are met, the user reports basic measurements to its serving BS which in its turn will decide whether a handover needs to happen. When a handover is triggered, the user gets assigned to the desired BS and the beam of that BS is properly aligned to the user. Perfect beam alignment is always assumed and achieved at zero cost. To make the simulation more realistic, it is assumed that beams are tracked and updated as function of mobility with varying update rates as will be explained later.
\subsection{Scenarios}
\par We consider a case study where a UAV is flying at a fixed speed of $14$~m/s (approx. $50$~km/h) with a fixed altitude above the city of Leuven. A set of random trajectories is generated where the UAV flies in a straight horizontal line in a random direction at two different altitudes of $40$~m and $150$~m. The same set of trajectories is used for every different set of parameters, this allows for a one-to-one comparison of the results. The baseline is the static scenario, where no beamtracking is enabled and where each sector antenna array is fixed towards its azimuth angle, tilted towards the ground at an angle of $7$ degrees. A X by 2 antenna array is used, with X adapted to the total number of antennas.
\par To show the impact of perfect beam alignment, we introduce some error in the alignment. The BS only periodically updates its beam alignment, which results in the serving beam not always being perfectly aligned. In the simulator, an update period of $0.1$~s is a realtime update period as the time step of the simulator is equal to $0.1$~s. The two antenna array sizes being compared are a 64-element array and a 256-element array. When comparing arrays of different sizes, square array configurations are considered. Different antenna array topologies are investigated for a 64-element and a 256-element antenna array.
\par To investigate the performance of different scenarios and topologies, we will look at the outage cost, which represents the fraction of the total travel time where the Signal-to-Noise (SNR) ratio is below the threshold of $-6$~dB, which is the minimum SNR required to provide a command-and-control link \cite{Colpaert2018}. An outage cost of one means the complete trajectory had an SNR lower than the threshold. Another performance parameter considered are the number of handovers per minute experienced. Expressing the handovers in this manner gives a good insight in the severity of this handover problem. All previously mentioned simulation parameters can be found in Table~\ref{tab:parameters}.

\begin{table}[bp]
\caption{Simulation parameters}
\begin{center}
\begin{tabular}{|c|c|}
\hline
\textbf{Parameter} & \textbf{Value}       \\
\hline
\hline
UAV heights, $h$       & $40$~m, $150$~m   \\
center frequency, $f$       & $26$ GHz      \\
signal bandwidth, $B$       & $400$ MHz      \\
transmit power, $p_{tx}$    & $18$ dB       \\
antenna element max gain, $G_0$ & $8$~dBi\\
antenna element 3dB beamwidth & 65~deg \\
noise density, $N_0$          & $-174$ dBm/Hz        \\
noise figure, $F$               & $9$ dB         \\
outage treshold & $-6$~dB \\
A3 treshold, $T_{A3}$ & $3$~dB\\
update periods & $0.1$~s,$0.2$~s,$0.5$~s\\
total number BSs & 20\\
total number sectors & 62\\
map area&$16~km^2$\\
number of trajectories        & $200$ \\
\hline
\end{tabular}
\label{tab:parameters}
\end{center}
\end{table}

\section{Results}
\par
\begin{figure}[t]
% \centerline{
% \includegraphics{handover_per_min}
% }
\centering
\begin{subfigure}[t]{0.5\linewidth}
\centerline{\includegraphics[width=\linewidth]{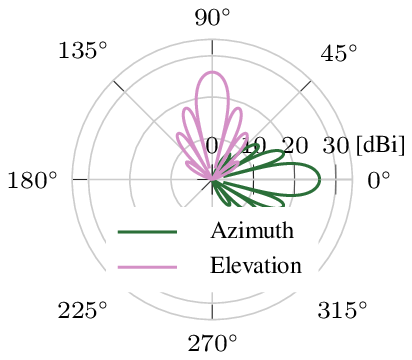}}
\caption{8 by 8 antenna array}
\end{subfigure}~
\begin{subfigure}[t]{0.5\linewidth}
\centerline{\includegraphics[width=\linewidth]{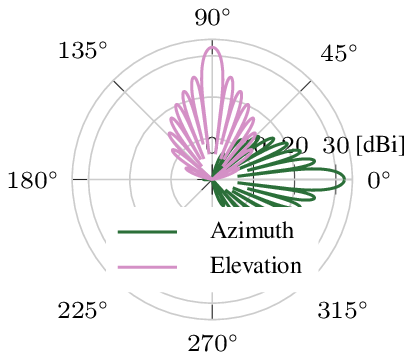}}
\caption{16 by 16 antenna array}
\end{subfigure}
\caption{The azimuth and elevation antenna patterns of an 8 by 8 and a 16 by 16 square antenna array}
\label{fig:antenna_increase}
\end{figure}

 First we shed some light on the effect of increasing the number of antenna elements in an array on the width of the beam when performing beamforming. We consider an array of 64 antenna elements and an array of 256 antenna elements, both in a square shape. The maximum gain of these antenna arrays are $26.1$~dBi and $32.1$~dBi, respectively. While the $3$dB beamwidths of the main beam are $18$~deg and $10$~deg, respectively. The resulting antenna array patterns can be seen in Figure~\ref{fig:antenna_increase}. Because the antenna array is square, the shape is the same in both the azimuth and elevation planes. The increase in number of antennas improves the maximum gain by $6$~dB (by quadrupling the antenna elements), but simultaneously reduces the beamwidth by $8$~degrees. Our simulations show that further quadrupling the antenna elements further reduces the beamwidth by $4$~degrees. While a narrow beamwidth is ideal in terms of minimizing interference to other users, it can become a hindrance when the beam becomes so narrow it is almost impossible to align, let alone track the user.
\par
\begin{figure}[t]
% \centerline{
% \includegraphics{handover_per_min}
% }
\centering
\begin{subfigure}[t]{0.5\linewidth}
\centerline{\includegraphics[width=\linewidth]{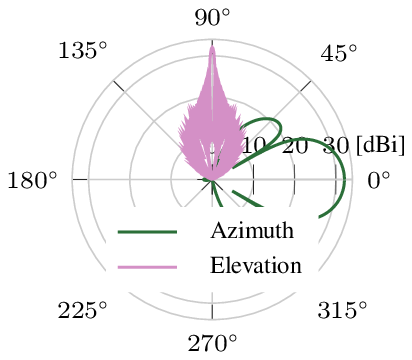}}
\caption{64 by 4 antenna array}
\end{subfigure}~
\begin{subfigure}[t]{0.5\linewidth}
\centerline{\includegraphics[width=\linewidth]{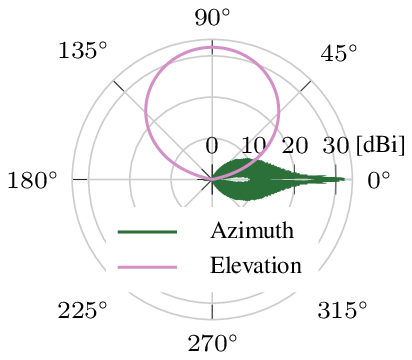}}
\caption{2 by 128 antenna array}
\end{subfigure}
\caption{The azimuth and elevation antenna patterns of an 64 by 4 and a 2 by 128 square antenna array}
\label{fig:diff_top}
\end{figure}

 Next, different configurations of a 256-element antenna array are considered. A rectangular array with more horizontal elements than vertical elements will give a more narrow beam and more spatial resolution in the horizontal plane and vice versa. A large spatial resolution is useful when the BS needs to serve for example multiple users in that specific plane, because it can spatially separate the users with a very narrow beam even when the users are near each other. However, when the user starts moving it will be more difficult to track the user in this plane because of the narrow beam.
% Less elements in a certain plane will generate a wider beam in that plane, resulting in more interference generated to nearby users in this plane \sofie{I would not mention this as you don't study it. Keep it to the point.}, but beamtracking with a wider beam leaves room for beam misalignment.
Keep in mind that in all configurations the maximum gain stays the same, thus this would allow for wide beams with large gains. The resulting antenna radiation patterns can be seen in Figure~\ref{fig:diff_top}. The first pattern (a) allows for very precise beamforming for when a user moves towards or away from the BS, while when the user moves in the horizontal direction with respect to the BS a wide beam is provided, allowing for larger beam alignment errors. The second pattern (b) behaves the opposite, practically generating vertical beam slices allowing misalignment in the vertical domain and precise beamsteering in the horizontal domain.
\par
\begin{figure}[tbp]
\centering
% \centerline{
% \includegraphics{handover_per_min}
% }
% \include{outage_time1}
\centerline{\includegraphics[width=1\linewidth]{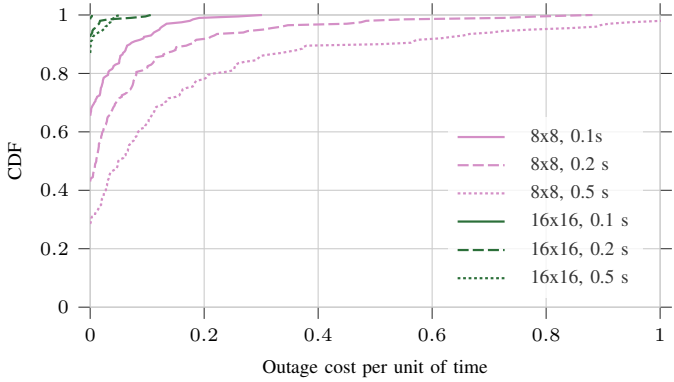}}
\caption{Outage cost for antenna array sizes of 64 and 256 antenna elements for different beam alignment intervals for a UAV flying at $40$~m above ground level.}
\label{fig:outage}
\end{figure}

 Following simulations compare the effect of increasing the number of antenna elements in an antenna array in terms of outage cost, which represents the fraction of time spend in a state of very low SNR below the outage threshold. As expected, the gain increases with increasing the number of antenna elements thus the SNR will improve, resulting in less coverage outages. The cumulative distribution function of each trajectory's outage cost can be seen in Figure~\ref{fig:outage}. As we can see, the coverage outage for the 16x16 array is significantly lower than that of the 8x8 array for drones flying at an altitude of $40$~m. However, when considering some beam misalignment, we see that the performance of both topologies drops, however, the impact is larger for smaller antenna arrays because of the already low SNR regime. We can conclude that for a drone flying at $40$~m, which should be a minimum altitude when crossing a city, the required number of antennas should be at least 256, especially when considering HD video streaming which will require an even higher SNR.
\par
\begin{figure}[t]
% \centerline{
% \includegraphics{handover_per_min}
% }
\centering
\begin{subfigure}[t]{\linewidth}
\caption{64-element antenna array}
\centerline{\includegraphics[width=\linewidth]{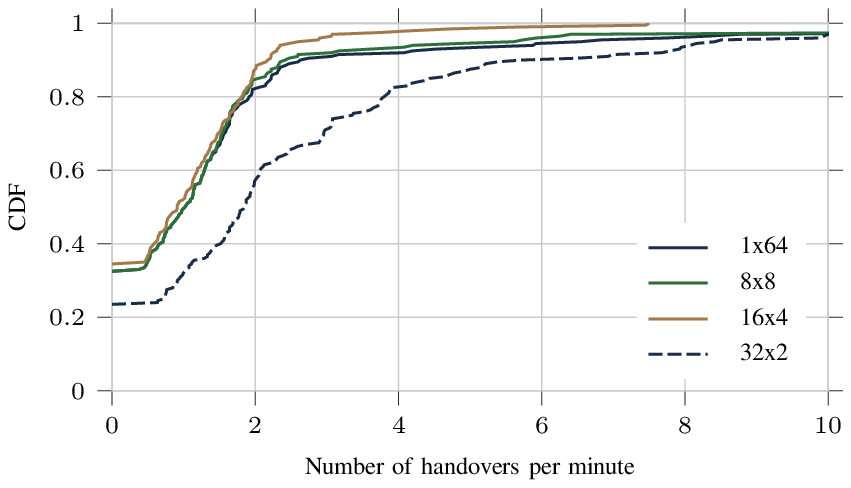}}
\end{subfigure}
\par\bigskip %
\begin{subfigure}[b]{\linewidth}
\caption{256-element antenna array}
\centerline{\includegraphics[width=\linewidth]{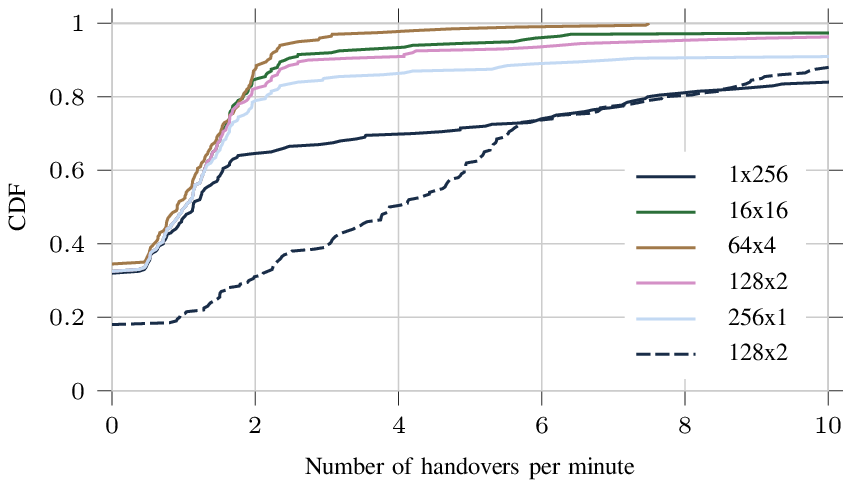}}
\end{subfigure}%
\caption{Cumulative distribution function plot of the number of handovers per minute for different antenna array topologies of a total of (a) 64 elements and (b) 256 elements at an altitude of $40$~m. The dashed line represents the baseline static scenario.}
\label{fig:handovers40}
\end{figure}

 Figure \ref{fig:handovers40} shows the number of handovers per minute for different antenna configurations of both a (a) 64-element and a (b) 256-element antenna array respectively at an altitude of $40$~m. The dashed line represents the baseline static scenario. A UAV in this scenario is mainly served by sidelobes resulting in larger numbers of handovers when compared to a beamsteering scenario. Additionally, increasing the number of antenna elements creates more sidelobes resulting in even worse handover rates as seen in \ref{fig:handovers40} (b). When considering beamsteering, for both numbers of antenna elements, the horizontal linear antenna array of 1 by 64 or 1 by 256 performs the worst of the set. This can be explained by the fact that from a BS perspective the relative movement of a UAV will mainly happen in the horizontal direction and a horizontal linear array generates narrow vertical plane shaped beam as shown in Figure \ref{fig:diff_top}. This shape results in the BS easily losing track of the user if it moves in the horizontal plane. The opposite is also true, a vertical linear array of 256 by 1 manages to track users moving in the horizontal plane, but lacks in its capabilities to track a user move towards/away from the BS. The square planar array should perform better in both planes perfectly, however users move in general more the horizontal plane. We could assume that a rectangular with slightly more antennas on the vertical axis should perform even better. Figure \ref{fig:handovers40} shows exactly this, an array of 16 by 4 or 64 by 4 elements slightly outperforms a default square array of the same number of antennas. We can also see that for larger antenna arrays the topology has more impact on the performance when compared to the results of the smaller antenna array.

% \begin{figure}[tbp]
% \centering
% % \centerline{
% % \includegraphics{handover_per_min}
% % }
% \include{handover_per_min40_2}
% \caption{Cumulative distribution function plot of the number of handovers per minute for different antenna array topologies of a total of 64 antenna elements at an altitude of $40$~m.}
% \label{fig:handovers40_2}
% \end{figure}
% \begin{figure}[tbp]
% \centering
% % \centerline{
% % \includegraphics{handover_per_min}
% % }
% \include{handover_per_min40}
% \caption{Cumulative distribution function plot of the number of handovers per minute for different antenna array topologies of a total of 256 antenna elements at an altitude of $40$~m.}
% \label{fig:handovers40}
% \end{figure}
\par
\begin{figure}[tbp]
\centering
% \centerline{
% \includegraphics{handover_per_min}
% }
% \include{handover_per_min150}
\centerline{\includegraphics[width=\linewidth]{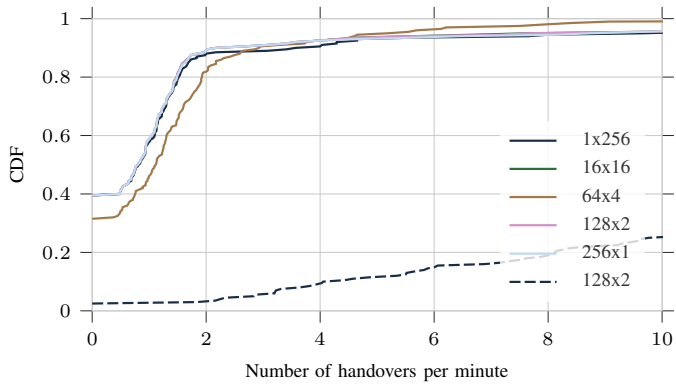}}
\caption{Cumulative distribution function plot of the number of handovers per minute for different antenna array topologies at an altitude of $150$~m. The dashed line represents the baseline static scenario.}
\label{fig:handovers150}
\end{figure}

 Figure~\ref{fig:handovers150} shows the same results as previous figure, but for UAVs flying at an altitude of $150$~m above ground level. We can conclude that the effect of the different antenna topologies is significantly less at an altitude of $150$~m. This can be explained by the fact that users have a larger distance towards the BS, thus the angular movement of the users is significantly less, which results in reduced beamtracking requirements.

\section{Conclusion}
In this paper, we suggest solving the excessive handover issue in UAV-enabled mmWave cellular networks by using beam forming and tracking on the BS side.
We analyzed different antenna array sizes and concluded that care should be taken when moving to larger antenna arrays.
The achieved array gains seem promising, but extra cost in antenna alignment should be taken into account as even the slightest misalignment will cause significant signal drops.
In all scenarios the beamforming approach reduced handover rate compared to the non-beamforming static sector approach. Next, we investigated the effect of using different antenna topologies (rectangular, vertical and horizontal rectangular arrays of various configurations) on the signal outage for the user as well as on the handover rate.
The general guideline that we suggest is to deploy slightly vertical rectangular antenna arrays (e.g., 64x4) if mobile aerial users should be served by the network.
These configurations are the most robust to the horizontal beam misalignment error (the most common one since UAVs typically move in xy-plane).
Additionally, we investigated the effect of flying altitude and concluded that UAVs flying at high altitudes generally experience less handovers as the relative mobility is lower, however, at a cost of lower signal strength. The opposite is true for the static scenario where the UAV sees even more sidelobes at high altitudes.
% \sofie{but other work suggested that flying higherr results in more handovers because of the sidelobes? can we refine this statement?}
\par
This work identified several problems that we plan to solve in future work.
We concluded that a UAV will experience on average at least one handover per minute going up to ten and more in worst case scenarios.
This means that the handover rates are definitely a problem for UAVs flying at a decent speed and that it will be difficult to achieve highly reliable links with high capacity.
However, we plan to suggest a handover procedure for UAVs specifically.
A possible approach could be smart handover skipping as in \cite{Arshad2016}.
% \sofie{but skipping might result in more outage, easy to simulate by Achiel.}
Moreover, Although the current deployment was sufficient to shed some light on the problems, future work might consider a more dense mmWave deployment, as it is planned in the future network deployments, and unearth even more problems as the cell size is further reduced.

\bibliography{biblio}{}
\bibliographystyle{IEEEtran}
\vspace{12pt}
\end{document}